\documentclass[conference]{IEEEtran}
\IEEEoverridecommandlockouts
\usepackage{cite}
\usepackage{amsmath,amssymb,amsfonts}
\usepackage{algorithmic}
\usepackage{graphicx}
\usepackage{textcomp}
\usepackage{xcolor}
\usepackage{tabularx}
\usepackage{subfigure}
\usepackage{multirow}
\usepackage{mathtools}
\usepackage{enumitem}
\usepackage{booktabs}
\usepackage{amsmath}
\usepackage{color}
\usepackage{bibentry}
\usepackage{bbold}

\def\BibTeX{{\rm B\kern-.05em{\sc i\kern-.025em b}\kern-.08em
    T\kern-.1667em\lower.7ex\hbox{E}\kern-.125emX}}
\begin{document}

\title{Pandemic Culture Wars: Partisan Differences in the Moral Language of COVID-19 Discussions}

\author{
Anonymous Authors
}

\author{
Ashwin Rao$^{*1,2}$,Siyi Guo$^{*1,2}$, Sze Yuh Nina Wang$^3$, Fred Morstatter$^1$, and Kristina Lerman$^1$\\
$^1$Information Sciences Institute,
$^2$University of Southern California,
$^3$University of Regina\\
\{mohanrao,siyiguo\}@usc.edu, syw332@uregina.ca,\{fredmors, lerman\}@isi.edu
}

\maketitle
\def\thefootnote{*}\footnotetext{These authors contributed equally to this work}\def\thefootnote{\arabic{footnote}}
\begin{abstract}
Effective response to pandemics requires coordinated adoption of mitigation measures, like masking and quarantines, to curb a virus's spread. However, as the COVID-19 pandemic demonstrated, political divisions can hinder consensus on the appropriate response. To better understand these divisions, our study examines a vast collection of COVID-19-related tweets.  We focus on five contentious issues: coronavirus origins, lockdowns, masking, education, and vaccines. We describe a weakly supervised method to identify issue-relevant tweets and employ state-of-the-art computational methods to analyze moral language and infer political ideology. We explore how partisanship and moral language shape conversations about these issues. Our findings reveal ideological differences in issue salience and moral language used by different groups. We find that conservatives use more negatively-valenced moral language than liberals 
and that political elites use moral rhetoric to a greater extent than non-elites across most issues. Examining the evolution and moralization on divisive issues can provide valuable insights into the dynamics of COVID-19 discussions and assist policymakers in better understanding the emergence of ideological divisions.
\end{abstract}


\maketitle

\section{Introduction}



The COVID-19 pandemic presented a major challenge to  society, testing its resilience and capacity to respond to a crisis. Slowing the rapidly spreading illness required coordinated adoption of non-pharmaceutical interventions recommended by public health experts, such as wearing a mask, staying home, social distancing, and getting vaccines once they became available. However, the response to the pandemic became politically polarized, with Republicans and Democrats disagreeing both on the gravity of the crisis and the appropriate measures to address it~\cite{pew2020partisan}. Some Republican leaders downplayed the severity of the virus and expressed skepticism about the effectiveness of measures like mask-wearing and social distancing, while some Democratic leaders advocated for more stringent measures like closing schools, businesses, and even parks and beaches. Policy differences in pandemic mitigation emerged at the state level as well. Republican-led states lifted restrictions on businesses and public gatherings earlier than Democratic-led states. When the COVID-19 vaccine was approved, many Democratic-led states introduced vaccine and mask mandates for returning to work and school, while some Republican-led states passed bills to ban such mandates. These political divisions hampered effective pandemic response, leading to more than a million deaths in US, one of the highest mortality rates among advanced economies~\cite{pew2020divide}.

To understand how the pandemic response became so polarized, we study a large corpus of pandemic-related tweets~\cite{chen2020tracking}, comprising over 270M tweets from 2.1M users around United States. We focus on five highly politicized and contentious wedge issues: coronavirus origins, lockdowns and business closures, masking, online education, and vaccines. These issues have been previously found to be  salient during the pandemic \cite{schaeffer2020lab, aidan2021lockdowns, rojas2020masks, luttrell2023advocating, chan2021moral, pierri2022online, rathje2022social}. We classify user ideology based on the partisanship of information sources they share and also describe a method for identifying issues raised in a tweet. These models enable us to label millions of tweets and users, thereby allowing us to study polarized discussions at scale.

We study online conversations about the pandemic from the lens of Moral Foundations Theory (MFT)~\cite{haidt2007morality}. This theory provides a framework for understanding how moral values shape people's political attitudes and behaviors. MFT proposes that individuals' values and judgments can be described by five moral foundations: care/harm, fairness/cheating, loyalty/betrayal, authority/subversion, and sanctity/degradation. Individuals and groups vary in how much emphasis they place on each foundation. While both liberals and conservatives consider the care and fairness to be important, conservatives are thought to endorse the latter three foundations to a greater extent than liberals ~\cite{graham2009liberals}. 
Studying online conversations from the perspective of MFT can provide insights into the ways in which partisans use moral appeals to articulate and defend their political positions. It can also help us understand how to frame messages so as to  appeal to the values of each group.

We organize our investigations around the following research questions:
\begin{enumerate}
    \item What COVID-19-related issues did liberals and conservatives discuss online?  

    \item How did attention to issues vary by ideology? Which issues were more salient to different groups and when? 

    \item How did liberals and conservatives differ in their use of moral language on different issues? How did the groups react to COVID-19-related events?
  
  
    \item How does moral language use differ between political elites and non-elites?
\end{enumerate}

We propose and evaluate a weakly-supervised method to identify issue-relevant tweets. This novel method extracts relevant phrases from publicly available Wikipedia pages and then uses them to detect issues in the text of tweets. In addition, we use state-of-the-art computational methods to detect moral foundations in tweets~\cite{guo2023data} and to infer the political ideology of users~\cite{rao2021political}. These models enable us to track partisan attention to polarized issues at scale from the start of the COVID-19 pandemic to November 2021. 

We uncover important differences in how liberals and conservatives discuss issues over time, as well as their use of moral language. In particular, we find that discussions of issues and reactions to events were negatively valenced, especially among conservatives. Liberals showed more sustained interest in the topic of education, while conservatives focused on the origins of COVID-19. We also unveil differences in the moral appeals of political elites and non-elites across the ideological spectrum. Understanding the moralization on divisive issues can provide valuable insights into the emergence of polarization and help public health experts better tailor messaging to the values and concerns of each group.

\section{Related Work}

\begin{table}
\small
\caption{Issue Detection: Wikipedia Articles}
\begin{tabular}{p{0.18\linewidth} p{0.77\linewidth}}
\toprule
    \textit{Issue} & \textit{Wikipedia Articles} \\
\midrule
  \textit{Origins} & Plandemic, Investigations into the origin of COVID-19, COVID-19 lab leak theory \\
\midrule
\textit{Lockdowns} & COVID-19 lockdowns, COVID-19 protests in the United States, U.S. federal government response to the COVID-19 pandemic, Protests against responses to the COVID-19 pandemic, Social impact of the COVID-19 pandemic in the United States, Stay-at-home order, Social distancing measures related to the COVID-19 pandemic, Quarantine \\
\midrule
  \textit{Masking} & Face masks during the COVID-19 pandemic, Face masks during the COVID-19 pandemic in the United States, Maskne, Anti-mask sentiment \\
\midrule
\textit{Education} &  Impact of the COVID-19 pandemic on education, Homeschooling during the COVID-19 pandemic, Impact of the COVID-19 pandemic on education in the United States \\
\midrule
  \textit{Vaccines} & COVID-19 vaccination in the United States, COVID-19 vaccine, COVID-19 vaccine hesitancy in the United States, COVID-19 vaccine misinformation and hesitancy, Deaths of anti-vaccine advocates from COVID-19, Herman Cain Award \\
\midrule
    \textit{Baseline} & COVID-19 pandemic, COVID-19 pandemic in the United States, Politics of the United States \\
\bottomrule
\end{tabular}
\label{tab:wiki_articles}
\end{table}

\subsubsection{Polarization}

American society has been divided by ``culture wars''~\cite{hunter1992culture}, with liberals and conservatives disagreeing on controversial issues (also known as wedge issues) like abortion, women's and LGBTQ+ rights~\cite{dimaggio1996have, evans2003have,abramowitz2008polarization}. 
This polarization was further exacerbated by the COVID-19 pandemic. Partisans disagreed on the virus's existence and severity, its origins, mitigation strategies like social distancing and masking, and vaccine mandates. Political ideology was found to explain differences in compliance with health guidelines and mandates ~\cite{gollwitzer2020partisan, grossman2020political}. A study of elite messaging on Twitter at the onset of the pandemic \cite{green2020elusive} found that Democratic elites discussed the pandemic more frequently and emphasized its threats to public health, while Republican elites focused on China and the pandemic's impact on businesses. 

These divisions extend beyond political elites: nearly 3 out of 10 Americans believed that COVID-19 was created in a lab, with 4 in 10 Republicans endorsing this claim \cite{schaeffer2020lab}. Government efforts to restrict the spread of COVID-19 through lockdowns were met with resistance in the form of protests and demonstrations, which were largely driven by conservatives: nearly $52\%$ of conservatives in the United States preferred to have fewer COVID-19 restrictions as opposed to only $7\%$ of liberals \cite{aidan2021lockdowns}. These divisions can have important behavioural consequences: following and interacting with right-leaning political elites, influencers and hyper-partisan media sources was found to be associated with vaccine hesitancy in the US \cite{pierri2022online, rathje2022social}. 

\subsubsection{Moral Foundations}

Differences in moral values may underlie these partisan divisions. We analyze these through the framework of moral foundations theory (MFT), which proposes five moral foundations of care, fairness, in-group loyalty, respect for authority, and purity/sanctity \cite{haidt2007morality}. These foundations can be further categorized into virtues (care, authority, fairness, loyalty \& purity) and vices (harm, subversion, cheating, betrayal \& degradation). A number of language analysis techniques have been used to quantify the moral foundations in text data (e.g. \cite{graham2012mfd, garten2016morality, guo2023data}), with past work suggesting that moral appeals may increase message diffusion \cite{brady2017emotion, wang2022diffusion}. 

To comprehend and mitigate issue-based political divisions, it is crucial to understand how individuals attach moral significance to their viewpoints \cite{koleva2012tracing}. Some past work has done this in the context of COVID-19. For example, \cite{diaz2022reactance} found that valuing the care foundation was correlated with compliance with COVID-19 public health recommendations. Similarly, \cite{chan2021moral} found that endorsing the fairness and care moral foundations predicted compliance with calls for staying-at-home, wearing masks, and social distancing, while purity predicted compliance with face masks and social distancing. Others have found that framing masking-related health messaging using ideology-matched moral arguments was effective for liberals, but not conservatives \cite{luttrell2023advocating}. Moral values can also be used to predict county-level vaccination rates: counties with residents who prioritize moral concerns about purity had lower vaccination rates while counties whose residents prioritized fairness and loyalty had higher vaccination rates \cite{reimer2022moral}.

Some initial work has assessed COVID-19 vaccine discourse on Twitter. \cite{borghouts2023understanding} found that liberal tweets about the COVID-19 vaccine expressed care, fairness, liberty\footnote{Note that some researchers include a sixth foundation, liberty/oppression.} and authority moral foundations more than conservative tweets, while oppression and harm were referenced more by conservatives. Similarly, \cite{pacheco2022holistic} found that care/harm was associated with pro-vaccine sentiment whereas, liberty/oppression was correlated with anti-vaccine attitudes. While vaccinations are a critical polarizing issue in the discussion of COVID-19, no work as of yet has explored differences in moral appeals across a broader range of contentious COVID-19 issues.

\section{Methods}

\subsection{Data}

We use a publicly available dataset \cite{chen2020tracking} consisting of 1.4B tweets about COVID-19 posted between January 21, 2020 and November 4, 2021. These tweets contained one or more COVID-19-related keywords, such as coronavirus, pandemic, and Wuhan, among others.  We utilized Carmen \cite{dredze2013carmen}, a geo-location identification technique for Twitter data to assign tweets to locations within US. This method leverages tweet metadata, including ``place'' and ``coordinates'' objects that encode location information, such as country, city, and geographic coordinates. Additionally, the technique uses mentions of locations in a user's bio on Twitter to infer their location. A manual review confirmed that this approach was effective in identifying a user's home state.  As a result, we were left with 270 million tweets generated by 2.1 million geo-located users in the United States.

\begin{figure*}[t]
    \subfigure[Origins]
    {\includegraphics[width=0.195\linewidth]{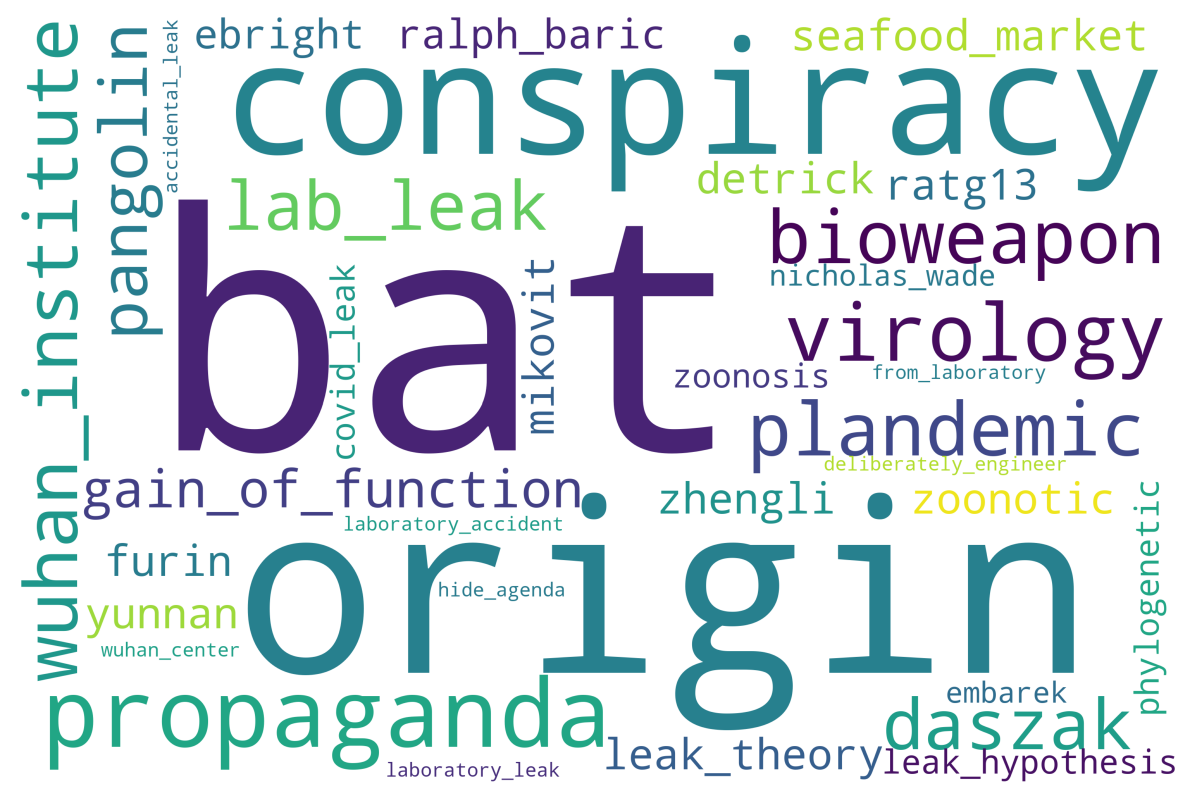}}
     \subfigure[Lockdowns]
    {\includegraphics[width=0.195\linewidth]{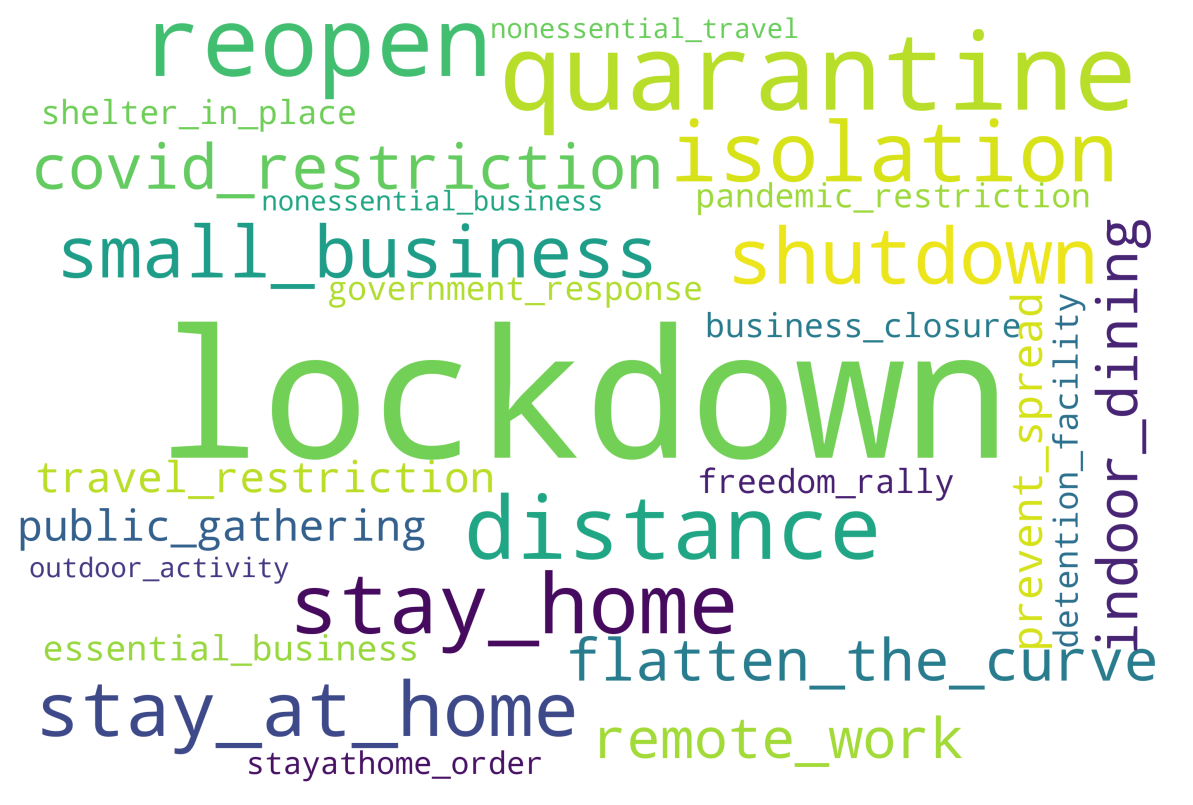}}
    \subfigure[Masking]
    {\includegraphics[width=0.195\linewidth]{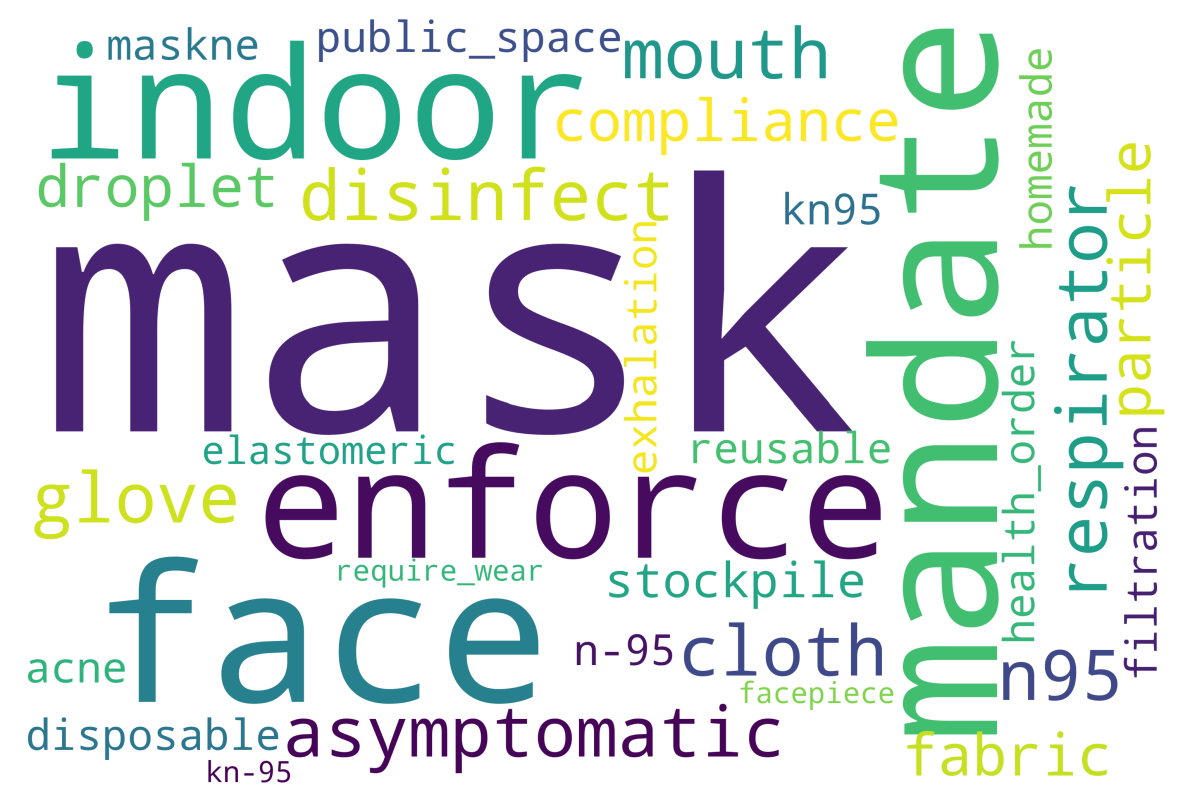}}
    \subfigure[Education]
    {\includegraphics[width=0.195\linewidth]{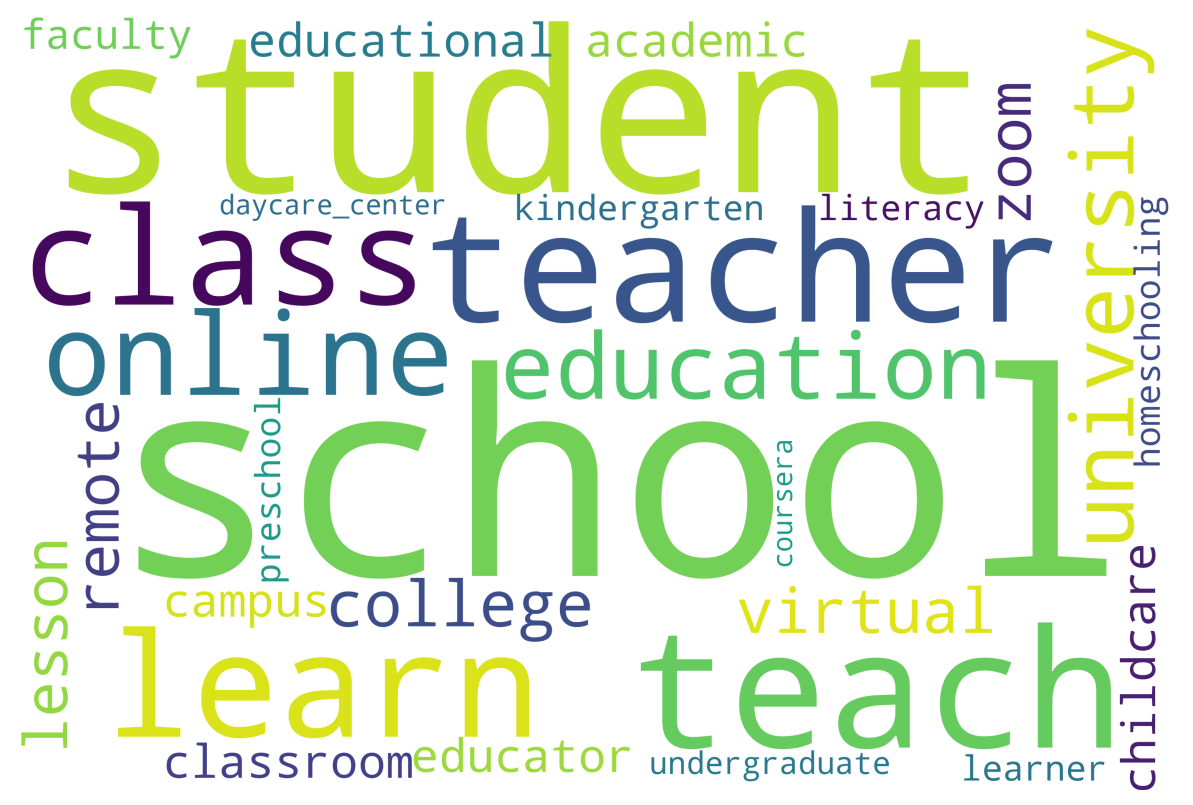}}
    \subfigure[Vaccines]
    {\includegraphics[width=0.195\linewidth]{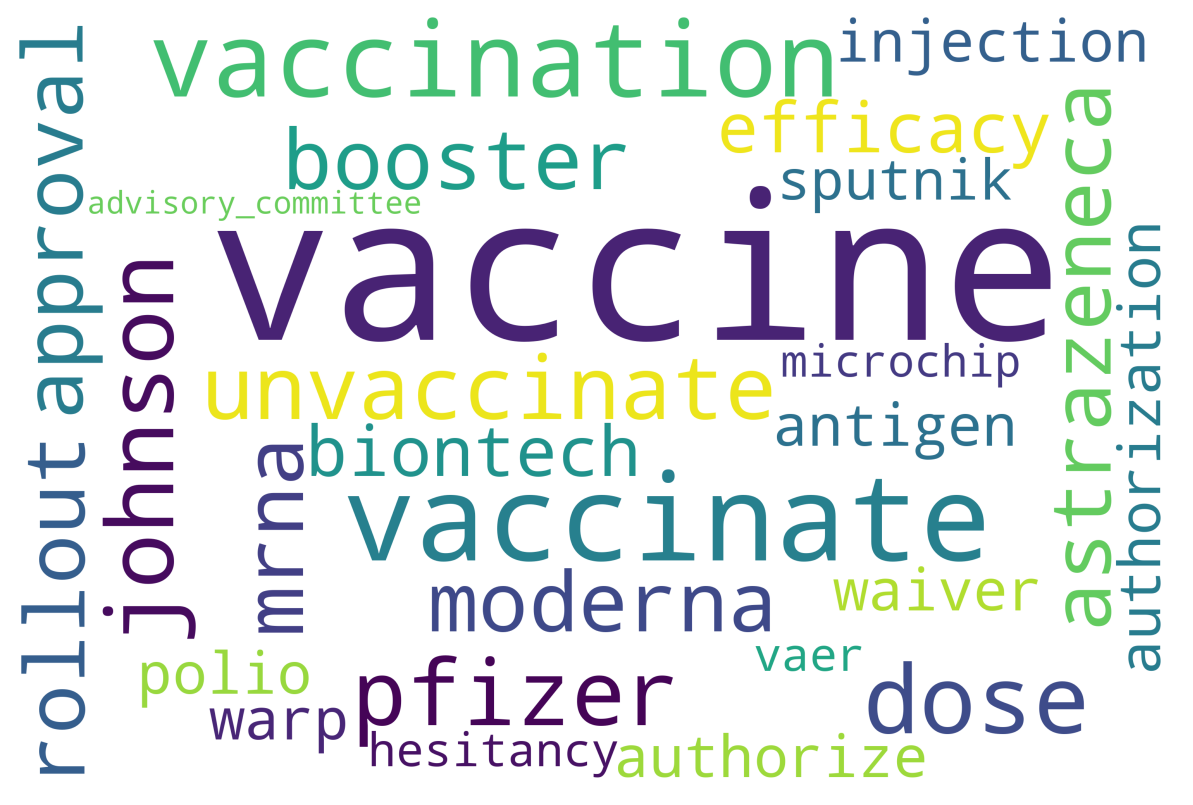}}
\caption{Wordclouds showing phrases identified from Wikipedia articles using SAGE for each issue. We show the top-50 most frequently occurring phrases in tweets. Bigrams and trigrams are connected by \_\_ for better visualization.}
\label{fig:word_clouds}
\end{figure*}

\subsection{Issue Detection}

\begin{table}
\caption{Sample tweets highlighting wedge issues in COVID-19 online discussions.}
\small
\begin{tabular}{p{0.17\linewidth} p{0.73\linewidth}}
\toprule
\textit{Issue} & \textit{Tweets} \\
\midrule
  \textit{Origins} & No matter what the Chinese Communist Party says, given the mounting evidence, the most likely \textcolor{red}{origins} for the China virus are the \textcolor{red}{Wuhan labs} studying bats and coronavirus. \\
  \midrule
\textit{Lockdowns} & This is a GREAT idea. We're all in this together. Take care of each other.  \textcolor{red}{\#StayHome} \#TakeItSeriously \#FlattenTheCurve \#COVID19 \\
\midrule
  \textit{Masking} & We’re in the middle of a pandemic and y’all are still coughing and sneezing without \textcolor{red}{covering your mouths}? Come on now. \\
  \midrule
\textit{Education} & More glimmers of hope as we “safely” move forward and open up Texas A\&M \textcolor{red}{University} while containing \#COVID19. \\
\midrule
 \textit{Vaccines} & You are joking right? Zero sympathy for \textcolor{red}{anti-vaxxers} who quit their jobs rather than get \textcolor{red}{vaccinated}. They put us all at risk and make the pandemic prolonged for the world. \\
\bottomrule
\end{tabular}
\label{tab:examples}
\end{table}

As the pandemic progressed, a set of polarizing issues such as origins of the coronavirus, lockdowns and masking mandates, school closures and online schooling and vaccines, emerged and grew to dominate conversations about COVID-19~\cite{schaeffer2020lab, aidan2021lockdowns, rojas2020masks, pierri2022online, rathje2022social}. 
Previous studies \cite{garimella2018political,brady2017emotion} relied on a small set of manually-selected keywords and phrases to harvest issue-relevant tweets.  However, there is a lack of data on the reliability of these methods, and a systematic framework to identify issue-relevant conversations is largely missing. While issue detection has analogues to topic-detection~\cite{blei2003latent, grootendorst2022bertopic}, one cannot control for the stochasticity of topics identified in traditional topic-detection approaches. Although \cite{card2015media} presented a corpus of several thousand news articles annotated with frames on traditional policy issues such as smoking, same-sex marriages, and immigration, there has not been an attempt to propose a framework for identifying COVID-19 issues in social media discourse. The lack of expertly annotated data on the issues we explore in this work prevents us from employing supervised learning techniques. 

We define the origins issue to encompass discussions surrounding the origins of the pandemic, including topics such as pangolins, gain of function research, wet-markets, and bats. The lockdown issue comprises content pertaining to early state and federal mitigation efforts, such as quarantines, stay-at-home orders, business closures, reopening, and calls for social distancing. The masking issue is defined by discussions on the use of face coverings, mask mandates, mask shortages, and anti-mask sentiment. Education-related content involves tweets about school closures, reopening of educational institutions, homeschooling, and online learning during the pandemic. The vaccines issue pertains to discussions about COVID-19 vaccines, vaccine mandates, anti-vaccine sentiment, and vaccine hesitancy in the US.

We describe a weakly-supervised method to harvest relevant keywords from Wikipedia pages discussing these issues (see Table \ref{tab:wiki_articles}). We use SAGE~\cite{eisenstein2011sparse} to identify distinctive keywords and phrases that are relevant to a specific issue. SAGE calculates the deviation in log-frequencies of words from a baseline lexical distribution.  
As a baseline, we use Wikipedia articles discussing general aspects of the pandemic and politics in the US (see \ref{tab:wiki_articles}). We concatenate these pages for each issue. Depending on how we tokenize the corpus, we can use SAGE to identify issue-relevant n-grams. In this study, we restrict analysis to unigrams, bigrams and trigrams. Using SAGE we then compare tokens from issue-relevant articles to baseline tokens to identify keywords and phrases that uniquely define each issue. We manually verify the keywords identified by SAGE to ensure precision and relevance. Word clouds in Fig.~\ref{fig:word_clouds} show the top 50 keywords/phrases extracted from Wikipedia articles for each issue. 
We assess the relevance of a tweet to a issue by the presence of these keywords and phrases. For example, terms such as `Wuhan labs' or `wet markets' would indicate a tweet is about the origins of the pandemic, while `cover your mouth', `N-95 masks' would indicate a masking-related tweet. Examples of tweets discussing these issues are shown in Table~\ref{tab:examples}.

\begin{table}
\centering
\caption{Inter-rater Agreement and Evaluation of Issue Detection.}
\begin{tabular}{p{0.17\linewidth} p{0.19\linewidth} p{0.15\linewidth} | p{0.07\linewidth} p{0.12\linewidth}}
\toprule
\textit{Issue} & \textit{Pairwise} & \textit{Multi-} & \textit{F1-} & \textit{Support}\\
& \textit{Cohen's $\kappa$} & \textit{annotator} & \textit{Score} & \\
& & \textit{Fleiss' $\kappa$} & & \\
\midrule
\textit{Origins} & $0.73 \pm 0.06$ & 0.45 & 0.51 & 25 \\
\midrule
\textit{Lockdowns} & $0.79 \pm 0.04$ & 0.47 & 0.90 & 108 \\
\midrule
\textit{Masking} & $0.92 \pm 0.04$ & 0.55  & 0.83 & 101\\
\midrule
\textit{Education} & $0.71 \pm 0.07$ & 0.50 & 0.74 & 54\\
\midrule
\textit{Vaccines} & $0.87 \pm 0.05$ & 0.53 & 0.92 & 84 \\
\bottomrule
\end{tabular}
\label{tab:issue_aggrement}
\end{table}

\paragraph{Validation}

We evaluate issue detection on a subset of tweets, which we draw at random with the constraint that there are at least five tweets for each combination of issues and moral foundation.
This gave us a test set of 784 tweets, which were then labeled by five trained annotators. Table~\ref{tab:issue_aggrement} shows that the mean Cohen's Kappa between each pair of annotators for different issues were all above 0.7. The Multi-annotator Fleiss' Kappa values are lower because it relies on agreement among all five annotators rather than pairs of annotators. Finally, we evaluate performance of issue detection method on this test set and show that F1 scores for most issues are above 0.7  (Table~\ref{tab:issue_aggrement}), indicating good model performance.



\subsection{Morality Detection}

We capture the moral sentiments in tweets based on the Moral Foundations Theory.  
We train our morality detection model on top of the transformer-based pretrained language model BERT \cite{devlin2018bert}. We train BERT with three Twitter datasets, including a manually annotated COVID dataset \cite{rojecki2021moral}, the Moral Foundation Twitter Corpus dataset with six different topics \cite{Hoover2020moral}, and a dataset of political tweets published by US congress members \cite{johnson-goldwasser-2018-classification}. 
Fusing an in-domain training set that is also about COVID-19, along with other datasets consisting of various topics, we are improving the model generalizability when applied to the target data \cite{guo2023data}. 

\subsection{Ideology Detection}
A number of methods identify the political ideology of social media users, 
with recent works \cite{le2019measuring, nikolov2020right, cinelli2021echo, rao2021political} leveraging URLs that users share in tweets. 
We use the method discussed in \cite{rao2021political} to estimate the ideology of individual users. This method relies on  Media Bias-Fact Check (MBFC) \cite{mbfc2023politics} to compute ideology scores for individuals. MBFC provides ideological leaning for over 6K Pay-Level Domains (PLDs) with leaning  categorized as Left/Hardline Liberal ($0$), Left-Center ($0.25$), Least-Biased/Center ($0.5$), Right-Center ($0.75$), Right/Hardline Conservative ($1$). Each individual's ideological leaning is estimated to be the weighted average of the leanings of the URLs they share. 



These scores are then binarized with a threshold of $\leq 0.4$ for liberals ($0$) and $\geq 0.6$ for conservatives ($+1$). The model described in \cite{rao2021political} then leverages a fastText embedding model pre-trained on Twitter data \cite{pgj2017unsup}, to generate embeddings for user's tweets over time. These embeddings serve as features to train a Logistic Regression classifier to predict the binarized ideological leanings for all users, including users who have not shared URLs from domains in the MBFC. We were able to identify ideology for 2.1M US users comprising of 1.6M liberal users and 500K conservatives.

\paragraph{Validation}
We assess the level of agreement between the ideology scores estimated by the methods proposed in \cite{rao2021political} and \cite{barbera2015birds}, which uses the follower graph to estimate user ideology. We discretize the continuous ideal point estimates in \cite{barbera2015birds} using a threshold of $< 0$ for liberals and $> 0$ for conservatives, which we respectively label as $0$ and $+1$. We identify approximately 35K users who appear in both datasets and calculate the F1-score between the two models. The resulting F1-score of $0.75$ and Jaccard overlap of $0.87$ indicates a high level of agreement between the two methods.

\begin{figure}[t]
    \includegraphics[width=0.99\linewidth]{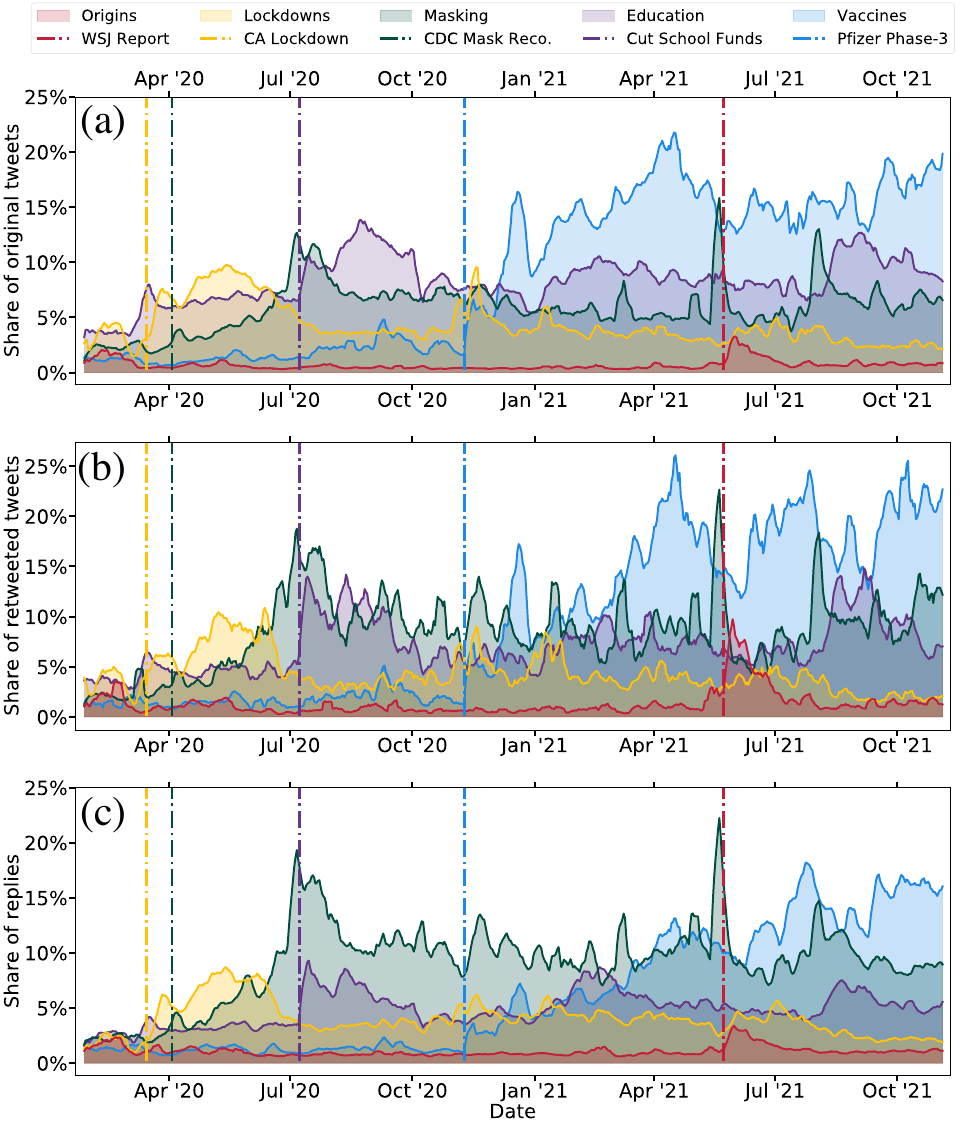}
\caption{Dynamics of attention to issues. Daily fraction of (a) original tweets, (b) retweets and (c) replies related to each issue. We use 7-day rolling average to reduce noise. Major events are marked with vertical lines of the same color as the issue and placed along the timeline.}
\label{fig:tweet_counts}
\end{figure}

\begin{figure}[tbh]
    \centering
    \includegraphics[width=\columnwidth]{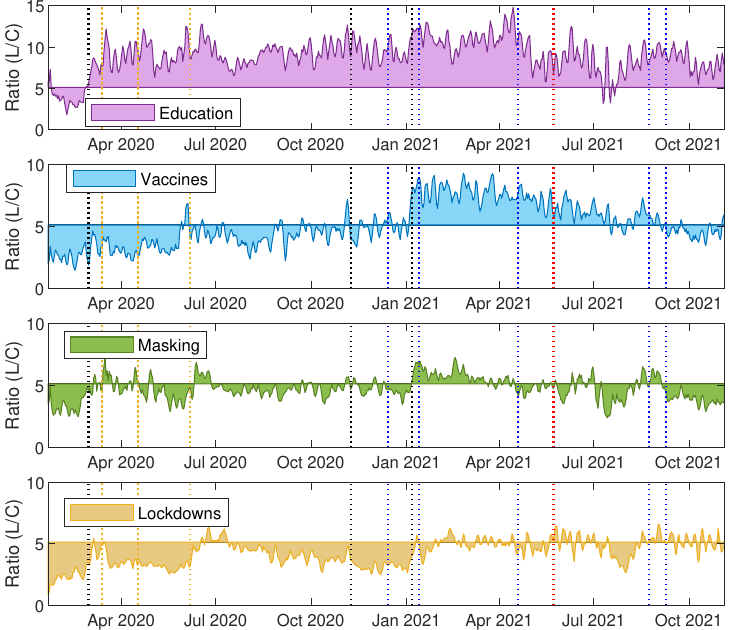}
    \caption{Partisan asymmetries in attention to issues. Each plot shows the ratio of the daily number of original tweets on a given issue posted by liberals and conservatives (smoothed using a three day window). Vertical lines correspond to major events during the time period: first US death 02/29/20, national emergency declaration 03/13/20, ``Liberate MI'' tweets 04/17/20, BLM protests 06/6/20, Biden defeats Trump 11/8/20, Covid vaccine rollout 12/14/20,  Jan 6 2021, Covid vaccine available to 65+ in CA 01/13/21, Covid vaccine available to 16+ 04/19/21, WSJ Covid lab leak story 05/23/21,  FDA approves first covid vaccine 08/23/21, federal vaccine mandate 09/9/21.}
    \label{fig:ratios}
\end{figure}


\begin{figure*}[t]
    \subfigure[Origins]
    {\includegraphics[width=0.195\linewidth]{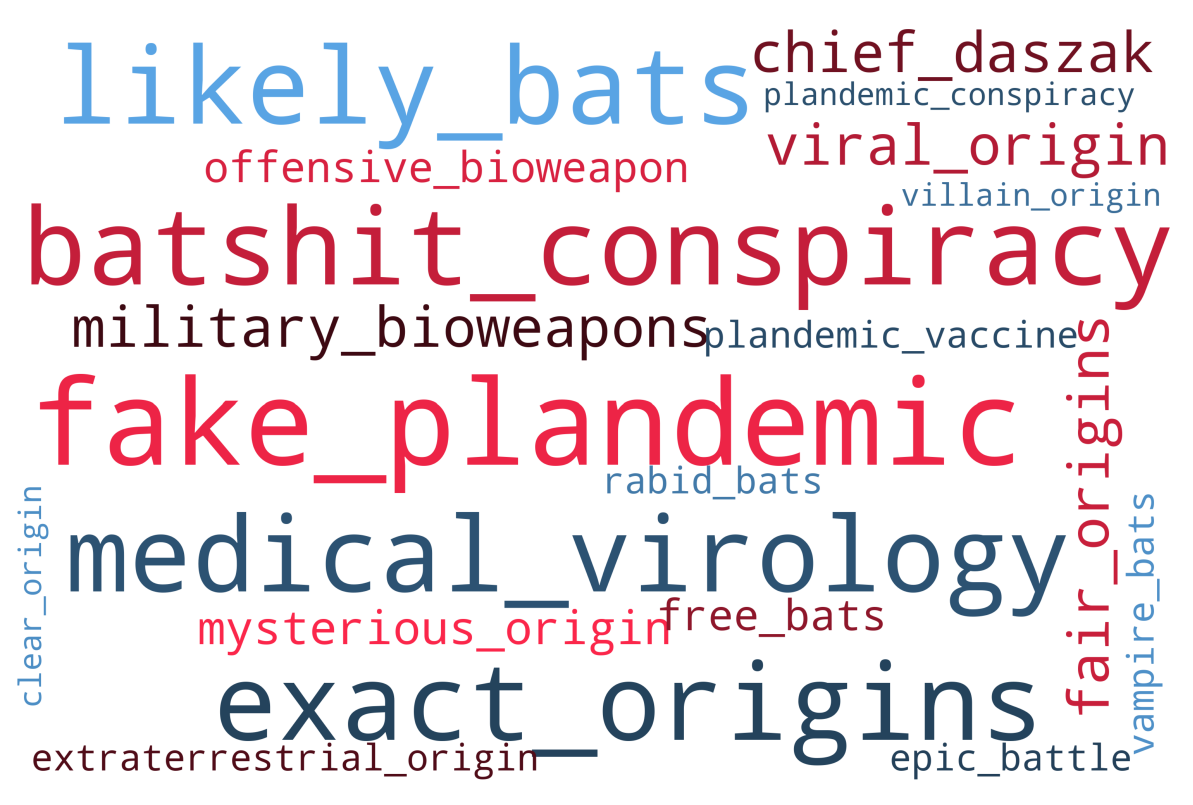}}
     \subfigure[Lockdowns]
    {\includegraphics[width=0.195\linewidth]{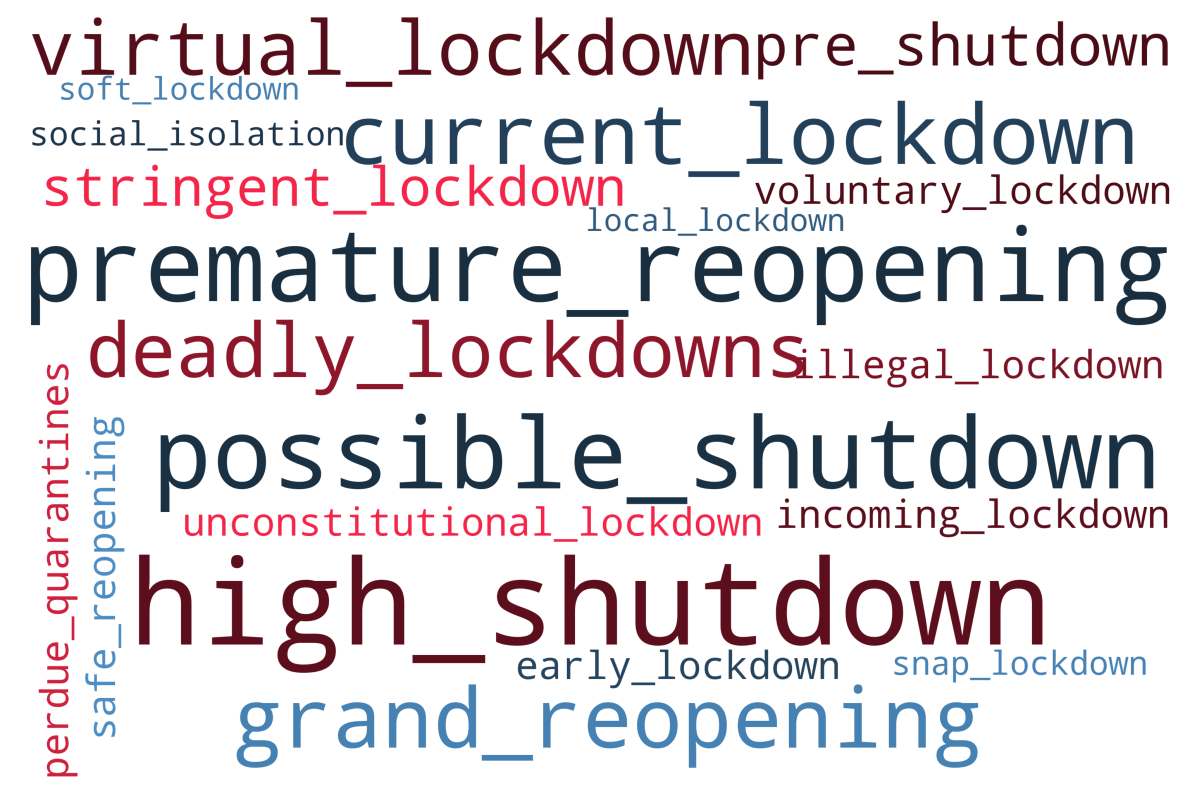}}
    \subfigure[Masking]
    {\includegraphics[width=0.195\linewidth]{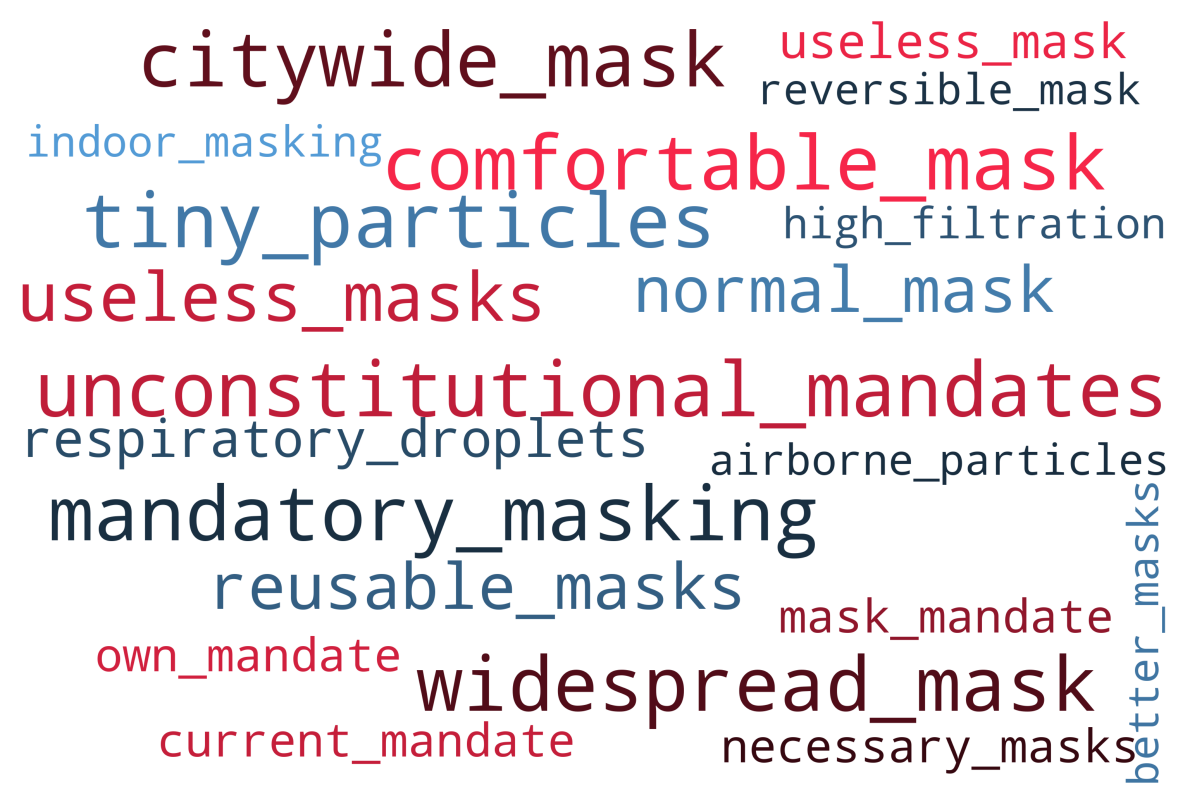}}
    \subfigure[Education]
    {\includegraphics[width=0.195\linewidth]{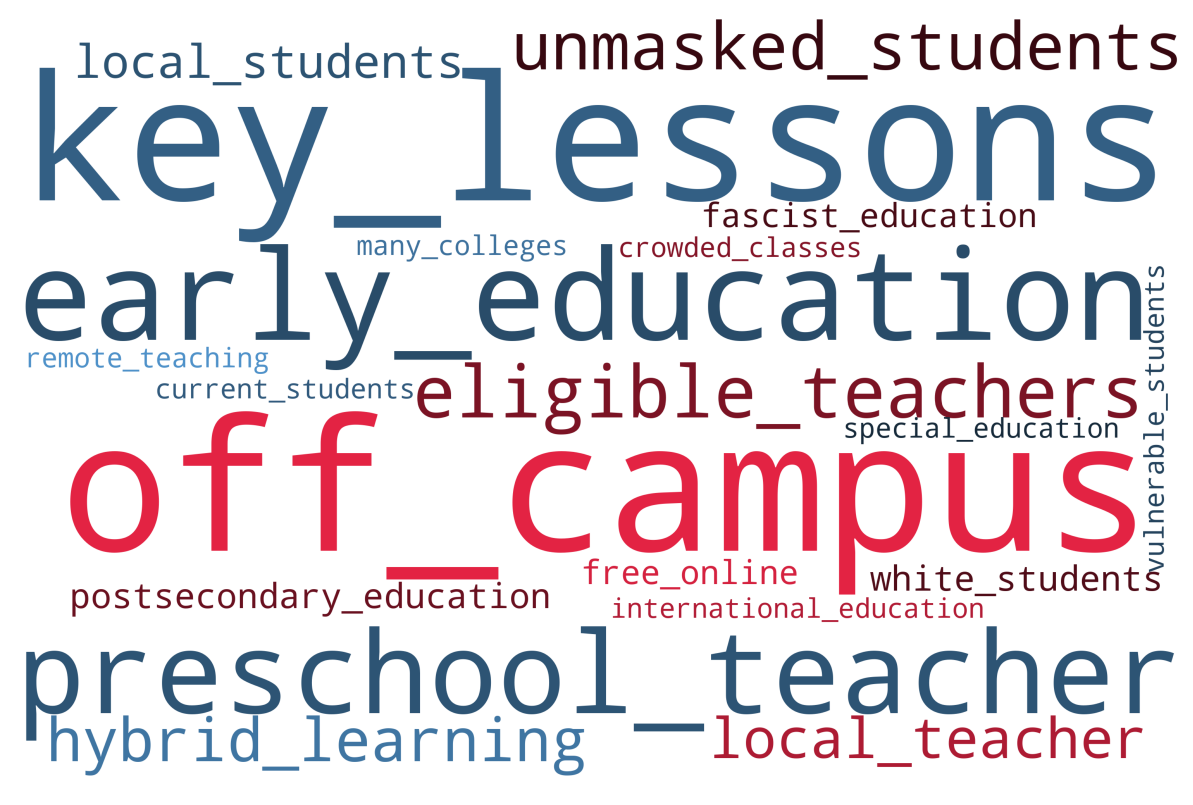}}
    \subfigure[Vaccines]
    {\includegraphics[width=0.195\linewidth]{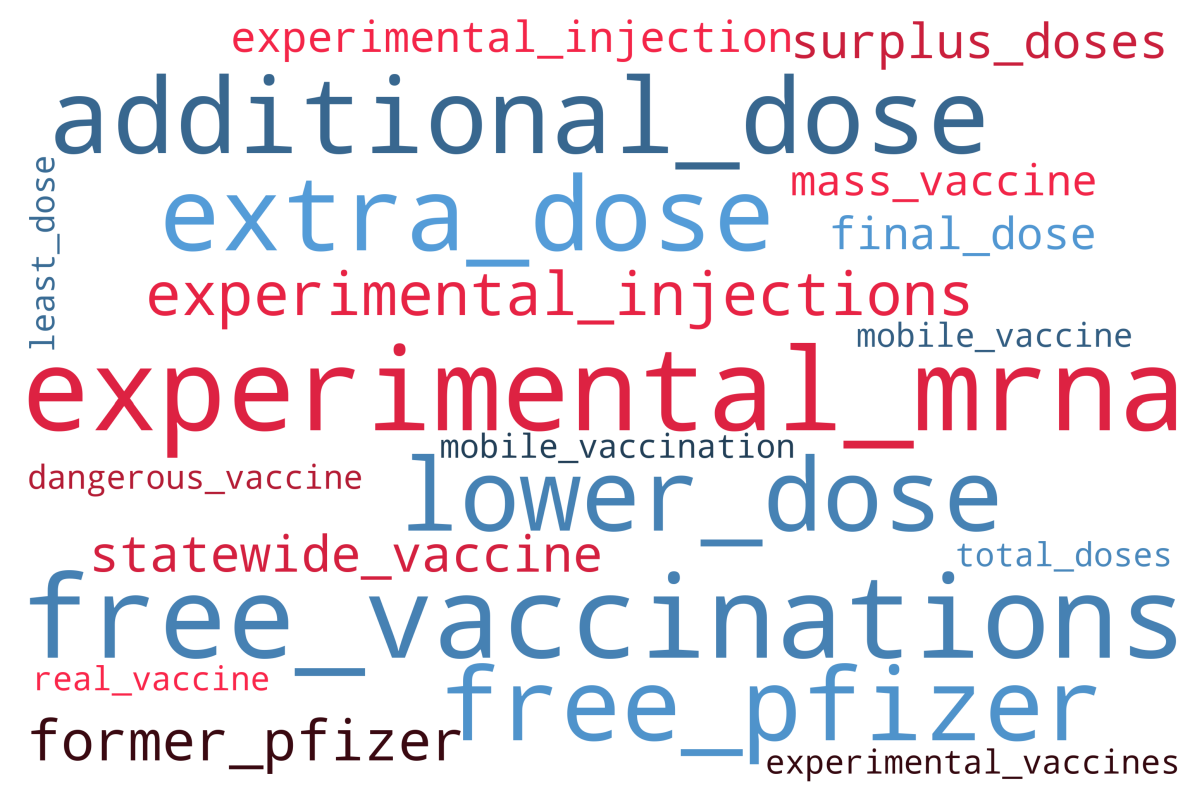}}
\caption{Partisan differences in framing each issue. Values show the log odds ratio of phrases used by liberals (blue) and conservatives (red) in issue-relevant tweets across the time period.}
\label{fig:frames_ideology}
\end{figure*}

\section{Results}

Figure~\ref{fig:tweet_counts} shows the daily share of tweets, retweets and replies discussing each issue, smoothed using a 7-day rolling window, with vertical lines marking major issue-related events. The five issues we monitored accounted for small share of all messages at the beginning of the pandemic, but grew to dominate all discussions. Lockdowns were among the first mitigation efforts carried out by state governments, with California the first state in the mainland US to issue a stay-at-home order on March 19, 2020. This date marks the start of an increase in lockdown related discussions. Although the peak in the masking discussions  
did not occur until the Black Lives Matter protests in June 2020, Fig.~\ref{fig:tweet_counts} indicates an early upward trend on April 3rd, 2020, when the CDC recommended face coverings to prevent infections. On July 8, 2020 President Trump promised to cut funding to schools that did not reopen for in-person instruction, which is associated with a spike in education related discussions. Similarly, Pfizer-BioNTech's announcement~\cite{pfizer2020vaxx} on November 9, 2020 that their 
COVID-19 vaccine  was highly effective in Phase 3 clinical trials created a spike in vaccine related tweets. A spike in the discussions of COVID-19 origins is seen on May 23, 2021 when the Wall Street Journal published an article citing an US Intelligence Report that three researchers from the Wuhan Institute of Virology had sought hospital care in November 2019 shortly before the outbreak \cite{wsj2021lab}. The story marked the first major report on the origins of the pandemic by a major news organization. These results provide a qualitative check of robustness of our issue detection, as well as insights into importance of issues at different times. Fig.~\ref{fig:tweet_counts} also shows differences in the engagement across issues. For example, tweets about masking seemed to garner more replies, suggesting that they may have generated more discussion or controversy. 

\begin{figure*}[t]
    \subfigure[Origins]
    {\includegraphics[width=0.19\linewidth]{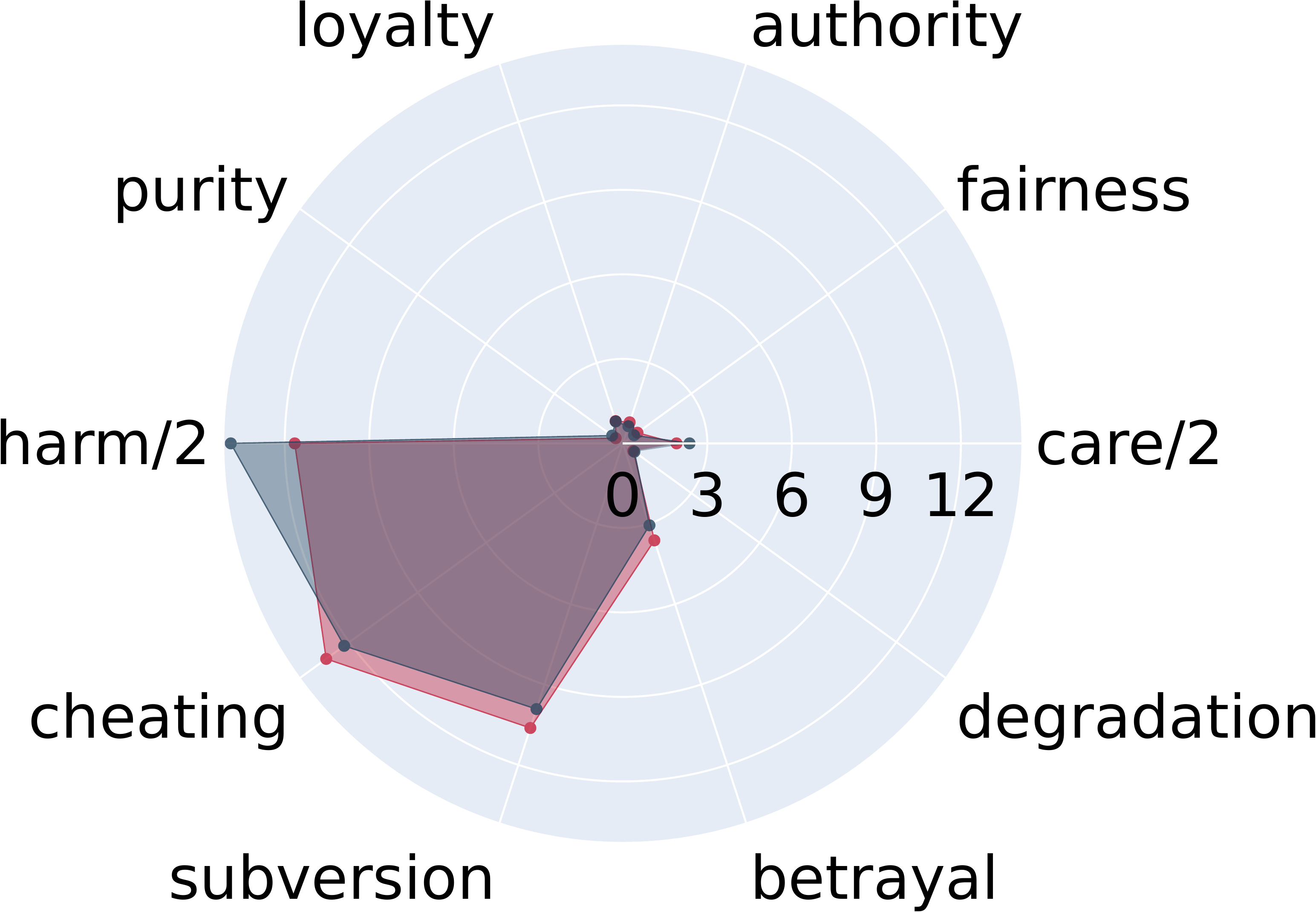}}
     \subfigure[Lockdowns]
    {\includegraphics[width=0.19\linewidth]{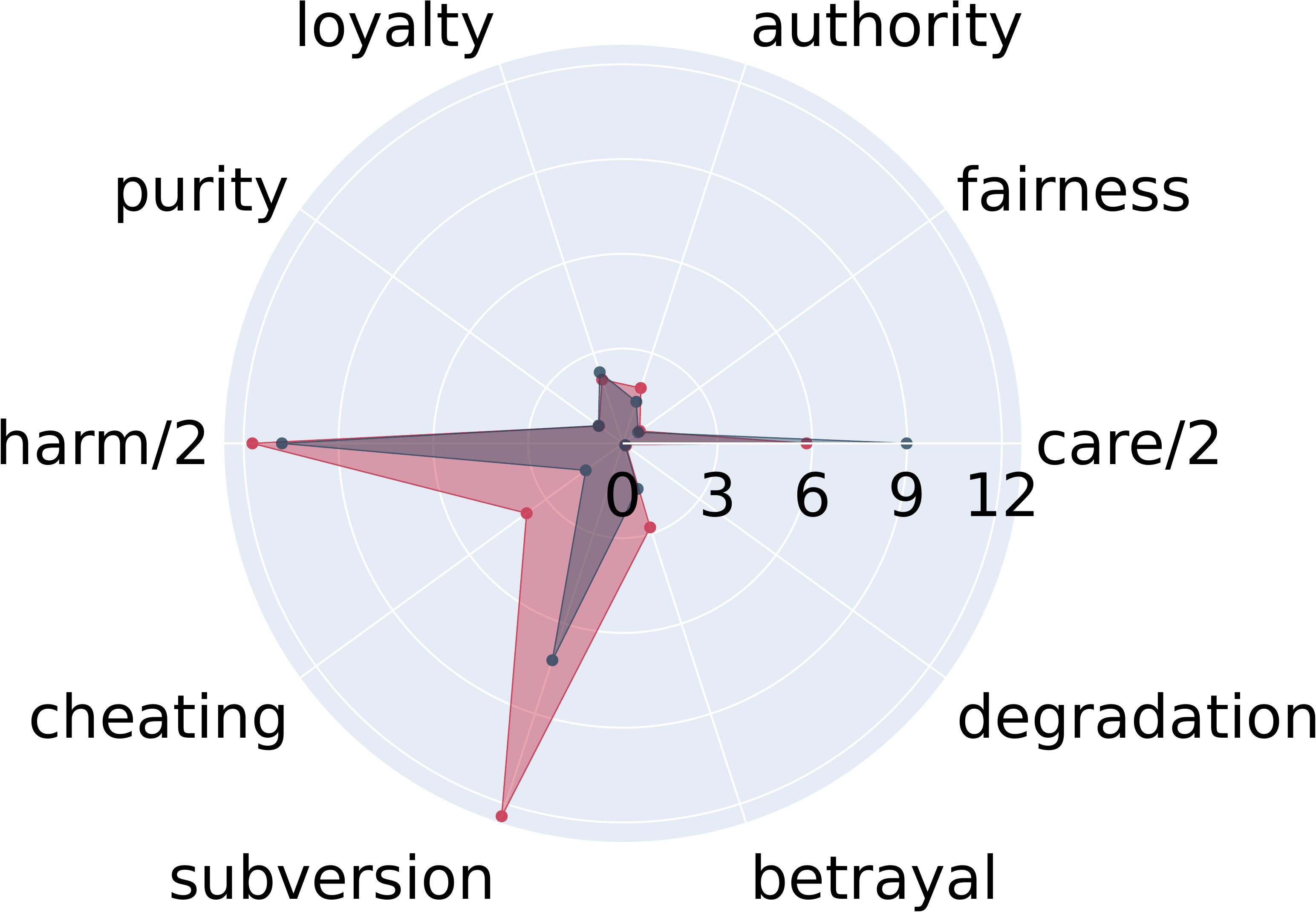}}
    \subfigure[Masking]
    {\includegraphics[width=0.19\linewidth]{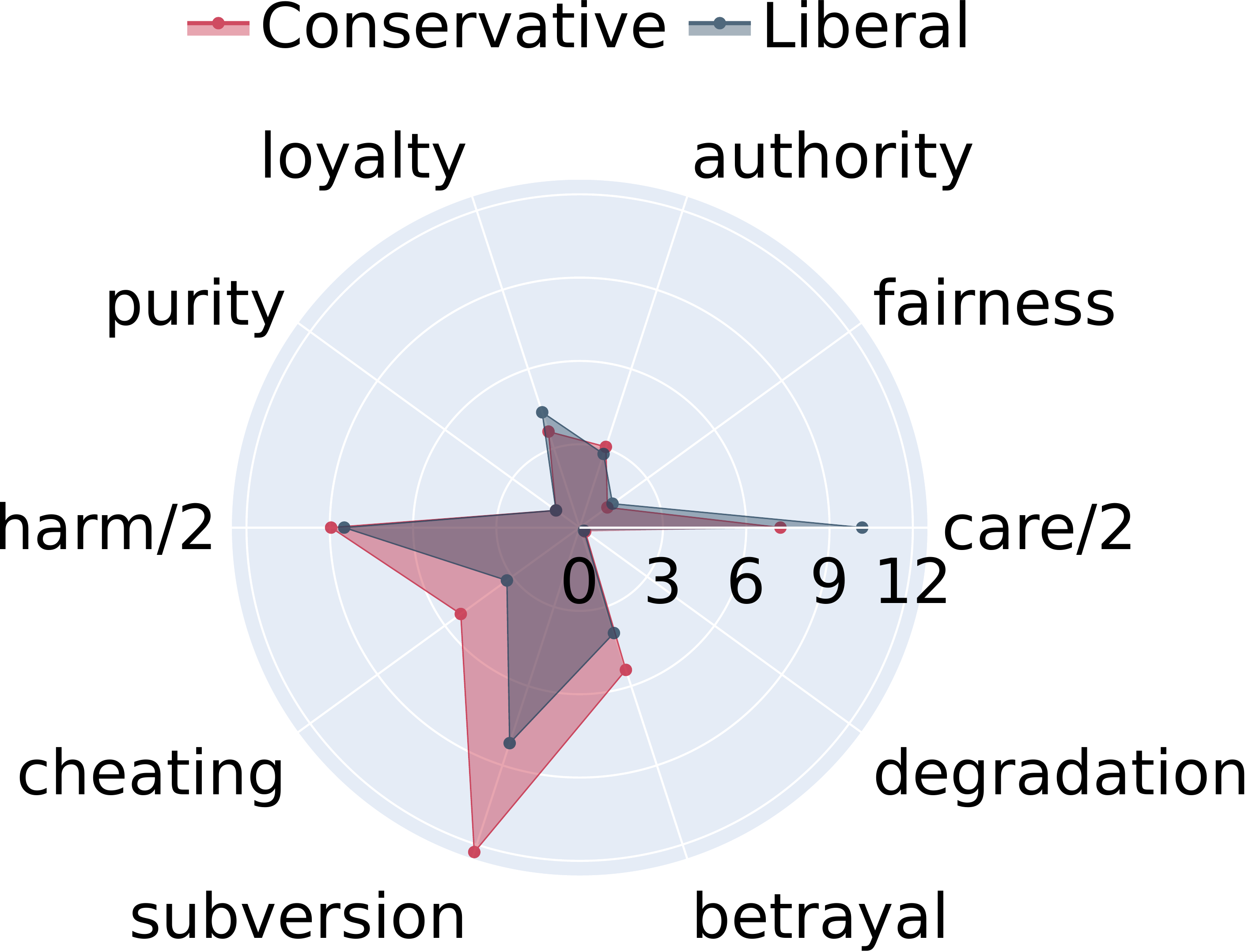}}
    \subfigure[Education]
    {\includegraphics[width=0.19\linewidth]{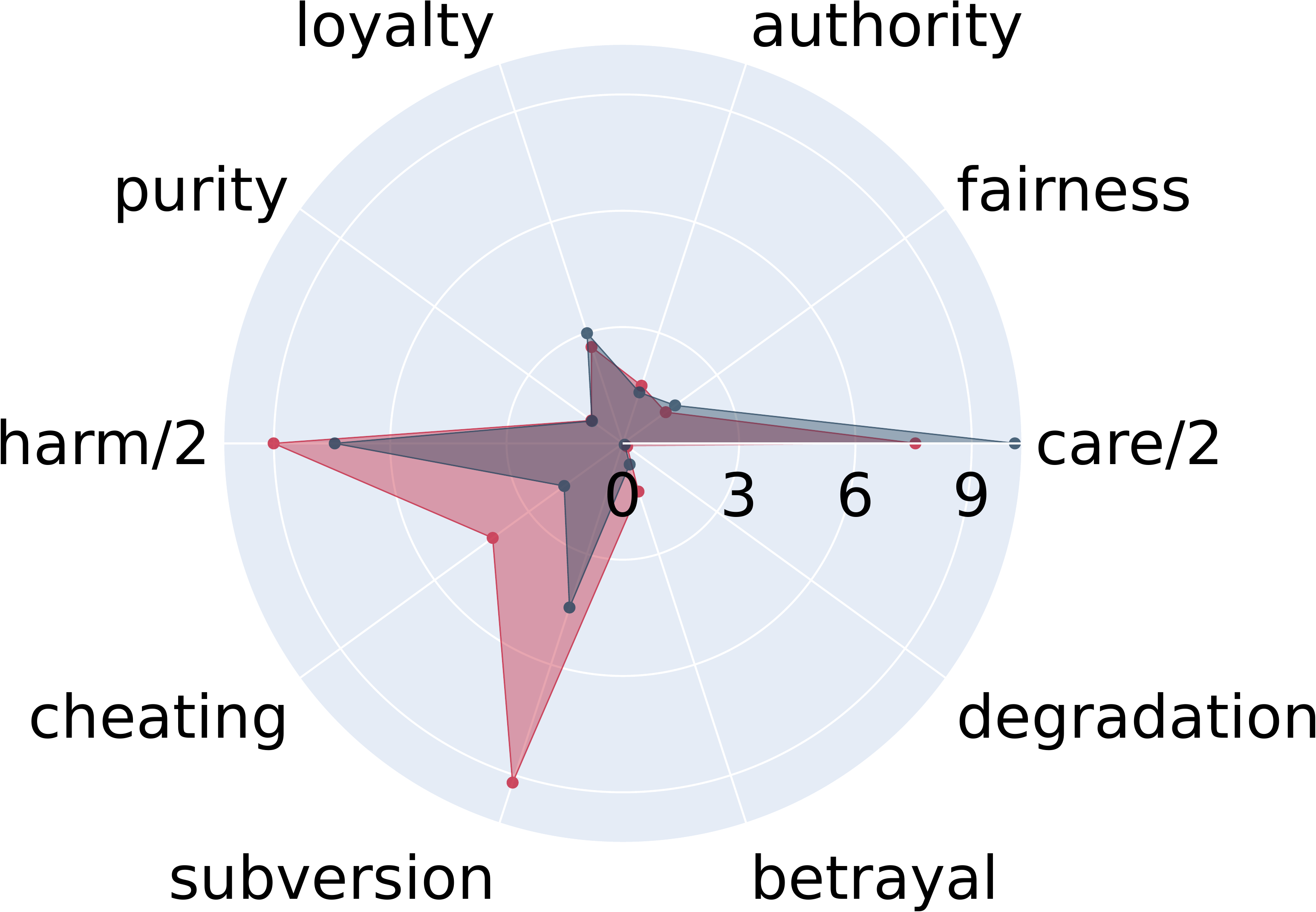}}
    \subfigure[Vaccines]
    {\includegraphics[width=0.19\linewidth]{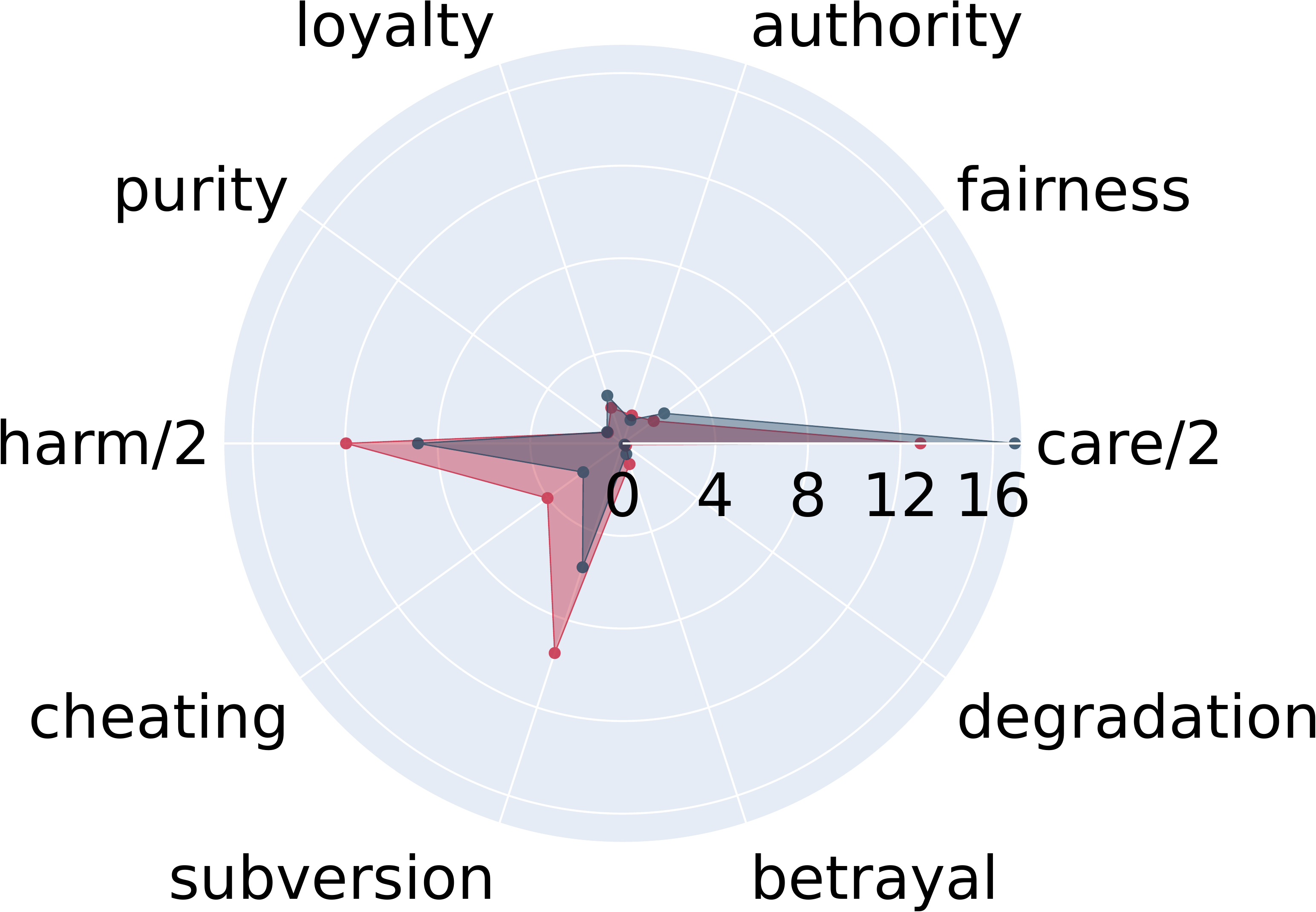}}
\caption{Partisan differences in moral language use along each issue. Values indicate percentage of tweets shared by liberals (blue) and conservatives (red) containing the specific moral foundation. Values for Care and Harm are downscaled by a factor of two for better visualization. }
\label{fig:morals_ideology}
\end{figure*}

\subsection{Partisan Differences in Attention to Issues}
We examine the relative prevalence of issues in the tweets of conservatives and liberals. The user population discussing the pandemic in our sample is ideologically unbalanced, with 5.143 times as many tweets by liberals as conservatives, on average; therefore, we cannot directly compare the volume of messages on an issue posted by each side. Moreover, the average imbalance hides a high degree of variability: the number of active liberals or conservatives may vary over time in response to events, making the issues more or less salient to each group. Using a measure like share (percentage) of tweets on an issue fails to capture the increased interest from a group. To address these challenges, we propose to use \textit{the ratio of the daily number of original tweets} to capture differences in partisan attention to issues.

Figure~\ref{fig:ratios} shows the time series of the daily tweet ratio $r(i,d) = {\tau_l(i,d)}/{\tau_c(i,d)}$, where $\tau_l(i,d)$ and $\tau_c(i,d)$ represent the number of original tweets posted by liberals and conservatives respectively about an issue $i$ on day $d$. 
Each plot shows the deviation of the ratio from its mean value. Overall, the education issue is of greater interest to liberals, with the mean ratio of liberal to conservative tweets being 8.951. In contrast, the issue of the origins of COVID-19 is of most interest to conservatives, with the mean ratio of 2.126. Masking and lockdowns are of slightly more interest to conservatives, with the mean ratios of 4.743 and 4.282 respectively. Different events make the issues more salient to liberals or conservatives, as captured by the positive or negative deviations from the mean. Specifically, the first US death due to COVID-19 was a turning point for liberals: their share of issue discussions increased. This is also confirmed by the event analysis, which shows more liberal reactions to this event. With respect to specific issues, although education is a liberal issue, conservatives show comparatively more sustained interest in it during the summer. The availability of COVID-19 vaccines at the end of 2020 (vertical lines show important events, including vaccine roll-out on December 14, 2020) led to more tweets from liberals  about the vaccine compared to conservatives. Likewise, national emergency declaration in March 2020, and anti-lockdown protests the following month brought sustained interest from conservatives to the lockdowns  issue. Following vaccine mandates for federal workers on September 9, 2021, the vaccines issue became a greater focus for conservatives, likely due to their outrage about the mandates. After Wall Street Journal's May 2021 report \cite{wsj2021lab}  claiming that COVID-19 originated at the Wuhan Institute of Virology increased conservative attention to the COVID-19 origins issue.
While we limit our discussion here to original tweets due to space constraints, we see similar trends in our analysis of retweets and replies.

To characterize differences in liberal and conservative tweets about issues, we identify adjectives that were used to modify issue-relevant phrases (anchors) extracted earlier. We leverage SpaCy's Dependency Parser to extract dependency relations of the form: $\textbf{XX}$-$amod \rightarrow ANC$ and $\textbf{XX}$–$amod \rightarrow YY \leftarrow amod$–$ANC$, where $XX$ refers to the extracted adjective and $ANC$ refers to an issue-relevant phrase. We then pair the extracted adjective with its corresponding anchor as a phrase of the form $XX$\_$ANC$ and compute the log-odds ratio of these phrases being expressed in liberal and conservative tweets. Wordclouds in Fig.~\ref{fig:frames_ideology} depict the top 10 most-likely used phrases in liberal (blue) and conservative (red) issue-discourse.

Conservatives use conspiratorial phrases such as ``batshit conspiracy'', ``fake plandemic'', and ``military bioweapon'' to discuss  origins of the pandemic. In contrast, liberals talk about   ``likely bats'' and ``medical virology''. When it comes to lockdowns, conservatives express negative sentiment, framing them as ``deadly'', ``unconstitutional'', and ``stringent''. Liberals, however, approach the topic cautiously, using phrases such as ``premature reopening'', ``safe reopening'', and ``soft lockdowns''. Similar patterns emerge in the discourse on masking, with conservatives deeming mask mandates as ``unconstitutional'' and ``useless'', while liberals focus on ``reusable masks'' and ``mandatory masking''. Concerning education, conservatives express distrust in mitigation strategies employed by educational institutions, using terms like ``fascist education'' and ``unmasked students'', while liberals highlight ``hybrid learning'', ``vulnerable students'', and ``early education.'' On the issue of vaccines, conservatives exhibit skepticism, referring to ``experimental mRNA'' and ``dangerous vaccines''. In contrast, liberals emphasize the benefits, discussing ``free vaccinations'' and ``mobile vaccines''.

\subsection{Partisan Differences in Moral Language} 

To study partisan differences in the use of moral language, we measure the share of issue-related tweets posted by each group that expresses each moral foundation. We examine moral valence by separately calculating the share  for vices and virtues. The proportion $\rho(m,i)$ of original tweets related to issue $i$ that express moral foundation vice or virtue $m$ is:
$
    \rho(m,i) = {\tau(m,i)}{\tau(i)},
$
\noindent where, $\tau(i)$ returns the number of original tweets on issue $i$.

Figure~\ref{fig:morals_ideology} shows partisan differences in the proportions $\rho$. Across issues, we find that care and harm the most prevalent moral virtues and vices, followed by authority and subversion, with fewer tweets expressing purity and degradation, despite the relevance of that moral foundation to the infection concerns. Conservatives express more moral vices than virtues in their tweets. There are other partisan differences in the use of moral language. For example, while both appealed to subversion, betrayal, and cheating when discussing the origins of COVID-19, liberals focus more on harm     (see Fig.~\ref{fig:morals_ideology}(a)). This could reflect greater concern among liberals about the potential harms of these speculations, as they led to a rise in anti-Asian hate crimes across the nation \cite{gover2020anti}.

Across lockdowns, masking, education, and vaccine issues (Figs.~\ref{fig:morals_ideology}(b)-(e)), conservatives expressed greater levels of harm and subversion than liberals, while liberals expressed more care. This may indicate greater support for mandates (e.g., school shutdowns and vaccine mandates) among liberals, and opposition by conservatives, consistent with partisan divides on shutdowns, social distancing, and masks \cite{pew2021covid}.

Overall, we find similarities and important differences in the moral language use by liberals and conservatives. Conservatives appear to make more negatively valenced moral appeals when it comes to lockdowns, masking, education, and vaccines, while liberals use greater care language. 

\subsection{Partisan Differences in Reactions to Events}
To understand how liberals and conservatives react to different events during the pandemic, we perform interrupted time series analysis. Our goal is to quantify both the changes in the use of moral language and the amount of discussions about different issues. First, we compute the time series of the partisan tweet ratios for each concern and moral foundation, $r(i,m,h) = {\tau_l(i,m,h)}/{\tau_c(i,m,h)}$, where $\tau_l(i,m,h)$ and $\tau_c(i,m,h)$ represent the number of original tweets posted by liberals and conservatives respectively about an issue $i$ and expressing a moral foundation $m$ in an hour $h$. We choose to use hourly data here to get more data points and therefore more robust regression results.

\begin{figure}[tbh]
    \centering
    \includegraphics[width=0.8\columnwidth]{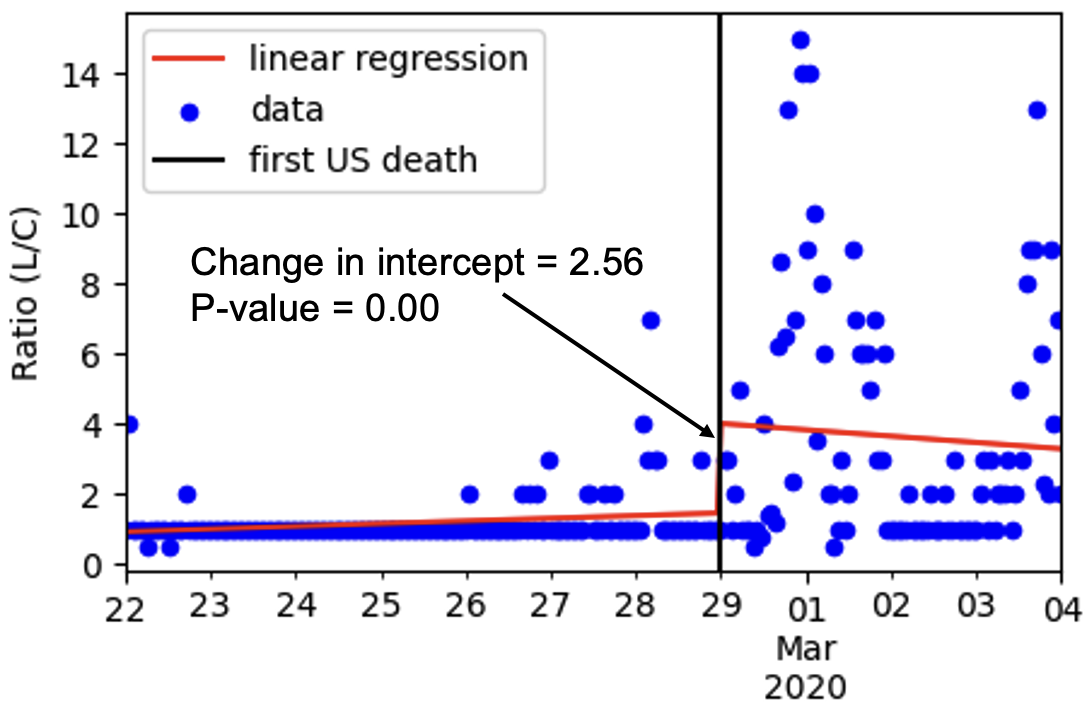}
    \caption{Interrupted time series analysis of response to the first US death (02/29/20). We carry out linear regression from seven days before the event to four days after it, using the time series of hourly ratio between the number of original tweets posted by liberals and conservatives respectively about an issue (specifically, \textit{education}) and expressing a moral concern (specifically, \textit{subversion}). The discontinuity, i.e., change in the intercept, shows more discussions expressing subversion about education from liberals than conservatives in response to the first US death.}
    \label{fig:its}
\end{figure}

We select four major events during the pandemic: (1) the first US death due to COVID-19 on 02/29/20, (2) the Black Lives Matter (BLM) protests on 06/06/20, (3) FDA approving the first COVID vaccine on 08/23/21, and (4) the federal vaccine mandate on 09/09/21. For each event, we select the time window from seven days before the event to four days after, as discussions on Twitter usually die down within that time period~\cite{leskovec2009meme}. We perform linear regression within this time window on each time series of the hourly tweet ratios:
$$
    r(i,m,h) = \beta_0 + \beta_1h + \beta_2\mathbb{1}_{after\ event} + \beta_3h * \mathbb{1}_{after\ event}
    \label{eq:event_analysis}
$$
\noindent where $h$ is \textit{hour}, and $\mathbb{1}_{after\ event}$ is 0 before the event and 1 after the event. To represent the change associated with an event, we use $\frac{\beta_3}{mean(r(i,m,h))}$, the coefficient for the interaction term normalized by the mean of this time series. Different issues have different mean ratios, so this normalization allows us to fairly compare changes across issues. Figure~\ref{fig:its} shows an example of such interrupted time series analysis with linear regression.

\begin{figure*}[tbh]
    \centering
    \includegraphics[width=0.8\linewidth]{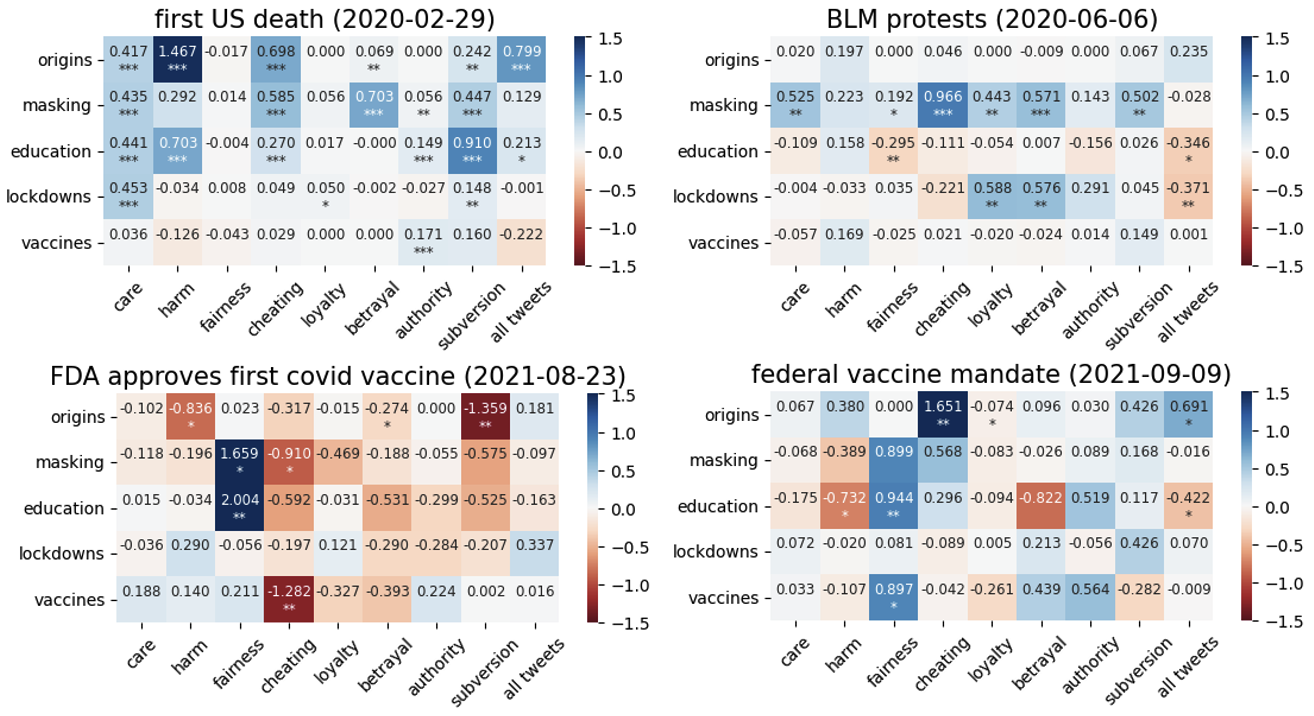}
    \caption{Different responses from Liberals and Conservatives to COVID-related events. The blue color (positive changes in L/C ratio after event) implies that there was a bigger reaction in liberals, and the red color implies that conservatives reacted more. Asterisks indicate significance values: * (p-value $< 0.05$), ** (p-value $< 0.01$), *** (p-value $< 0.001$), and no asterisk indicates p-value $\geq 0.05$.}
    \label{fig:event_analysis}
\end{figure*}

Figure~\ref{fig:event_analysis} shows how liberals and conservatives reacted differently to different events. The color of the cell relatively change in the discussions of the specific issue and moral framing: $\frac{\beta_3}{mean(r(i,m,h))}$. A positive value, colored blue, means that there was relatively more engagement from liberals in response to this event, and a negative value, in red, means that there was more conservative engagement. We see that liberals and conservatives had distinct moral reactions to different events. To begin with, the first US death attributed to COVID-19 on 02/29/20 elicited more reactions from liberals. This implies that liberals were paying more attention to the COVID-19 disease at this early stage of the pandemic. During the Black Lives Matter (BLM) protests on 06/06/20, conservative engagement on the issues of education and lockdowns increased overall. However, reliance of the cheating, loyalty/betrayal and subversion moral foundations on the issues of masking and lockdowns significantly increases among liberals. This suggests that liberals expressed more concerns about COVID safety during mass gatherings. More interestingly, the FDA approval of the first COVID vaccine exhibited diverging moral reactions. While, liberals focused on fairness more, the conservative reactions to this event were more negatively valenced, expressing more cheating, betrayal and subversion about different issues, suggesting a more pessimistic view of the COVID vaccines. On another vaccine-related event, the federal vaccine mandate, liberals consistently had more discussions about fairness, especially when talking about the issues of education and vaccines. These results show us the different moral reactions in the two partisan groups towards important events and agendas along the pandemic timeline.

\subsection{Differences in Moral Language of Elites and Non-Elites}

\begin{figure}[ht]
    \includegraphics[width=0.9\linewidth]{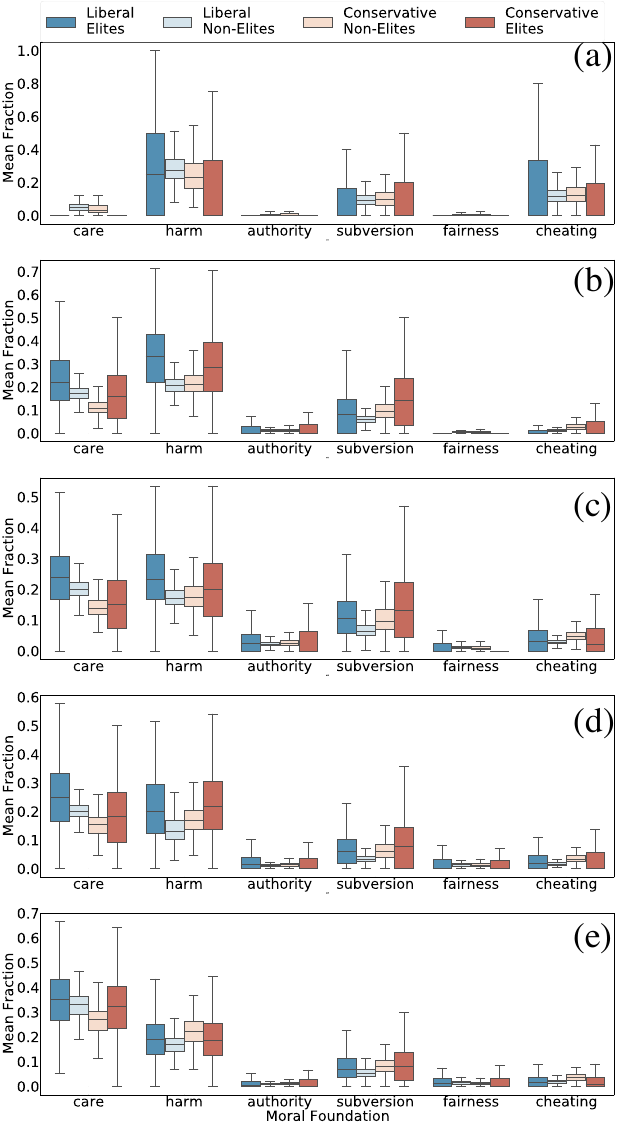}
\caption{Moral language differences between elites and non-elites. Boxplot shows the differences in the share of daily tweets made by each group on the issues: (a) origins, (b) lockdowns,(c) masking, (d) education and (e) vaccines.}
\label{fig:moral_elite}
\end{figure}

Past research has shown that moral language used by political elites~\cite{clifford2013words,wang2021moral} in the US can influence the opinions and moral reasoning of non-elites \cite{clayton2021elite}. In this work, we compare the moral language used by elites and non-elites. 
Political elites in                                      US include prominent members of Congress, politicians, journalists and media pundits. Previous studies \cite{wojcieszak2022most,shugars2021pandemics} have curated extensive lists of Twitter handles for over 4K US political elites. To ensure a fair comparison with non-elites, we randomly sample non-elite original tweets such that the number matches the number of tweets from elites on a given day. The random sampling also ensures that tweets from non-elites with higher overall activity are selected. We then calculate the daily proportion of issue-related tweets that express a certain moral foundation. Specifically, we calculate the proportion $\rho$ of original tweets relevant to issue $i$ that express moral foundation $m$ on day $d$: $\rho(m,i,d) = \frac{\tau(m,i,d)}{\tau(i,d)}$ where, $\tau$ returns the number of original tweets. We calculate this  separately for elites and non-elites. To account for variation, we bootstrap non-elite tweets 100 times.

Figure~\ref{fig:moral_elite}((a)-(e)) 
compares moral language use by elites and non-elites across issues. Boxplots show the distribution of the daily share of tweets in each category.
We do not show loyalty/betrayal and purity/degradation foundations, as they were not substantially expressed by elites. 
The lower variance for non-elites is an attribute of bootstrapping. We rely on the non-parametric Mann-Whitney U Test to test the significance of our comparisons. All differences are assessed for significance at $p<0.001$ unless otherwise specified.

On the issue of origins (Fig.~\ref{fig:moral_elite}(a)), non-elites express more care, subversion and cheating than elites.
Conservative non-elites use more harm language than conservative elites, a difference that is not significant for liberals. Lockdowns, masking, and education tweets all show a pattern of elites using more moral language than non-elites across most moral foundations, with liberal elites and non-elites tending to refer more to care (Fig.~\ref{fig:moral_elite}(b-d)). 



Discussions of vaccines deviate somewhat from these trends
(Fig.~\ref{fig:moral_elite}(e)), with conservative non-elites expressing more harm than both liberals and conservative elites. Conservatives  (elites and non-elites) also express more subversion than liberals.

\section{Conclusions}

The COVID-19 pandemic presented significant challenges to society both within and beyond the sphere of public health. The ensuing response to the pandemic quickly became polarized, with liberals and conservatives disagreeing on the severity of the pandemic and appropriate measures to address it. Sharp differences in issue positions were observed in discussions relating to the origins of the pandemic, lockdowns and business closures, masking mandates, disruptions to education, and vaccines. Previous studies \cite{schaeffer2020lab, aidan2021lockdowns, rojas2020masks, luttrell2023advocating, chan2021moral, pierri2022online, rathje2022social} attest to the salience of these issues during the pandemic. We use a massive corpus of over 200M tweets~\cite{chen2020tracking} to assess these ideological divisions. Our findings suggest that not only do conservatives and liberals differ in which issues they discuss and how they frame these issues, but also how they moralize these issues. We also demonstrate that a weakly-supervised classifier can be used to accurately identify issue relevant tweets.

While liberals dominated the discussion on education, tweets about COVID-19 origins and lockdowns were more prevalent among conservative. Tweets about the vaccine were predominantly made by liberals once the vaccine became available, but switched to being more important to conservatives after federal vaccine mandate were announced. We also found strong differences in moral language among liberals and conservatives. We found that tweets expressing the care/harm and authority/subversion foundations were most persistent, suggesting that these values resonated with both groups. This contrasts with past results suggesting that authority/subversion is primarily considered an important moral dimension by conservatives \cite{graham2009liberals}, and supports the idea endorsing the moral foundation of authority may be largely dependent on $which$ authorities are being discussed, and what ideological stereotypes are associated with those authorities \cite{voelkel2019mft}. 

Our analysis revealed further differences in the moral language of partisans. While conservatives relied on negatively-valenced moral language, like harm and subversion, liberals primarily invoked care. The endorsement of negatively-valenced morality in conservatives is especially reflected in their moral reactions to COVID-related events, such as the FDA approving the first vaccine. This has important implications for message tailoring and public policy: establishing trust in public health institutions is critical to increasing compliance with public health measures, as is addressing moral concerns about these measures. We also found that political elites in the US tended to use more moral language compared to non-elites, regardless of ideology. Given the greater influence that elites have, this may lead to outsize perceptions of the extent of political polarization on these issues, particularly as negative tweets by elites tend to spread particularly widely~\cite{hwang2014onlinedebate}. In sum, our findings suggest that differences in the expressions of moral values drove ideological divisions on different issues, in response to different events, fueling polarization. Overall, our findings suggest that polarization is reflected in 1) the issues that partisans focus on, 2) the way that they frame these issues, and 3) the moral values they apply to issues.

However, it is worth noting that our methods have several limitations. We infer geo-location of tweets based on tweet metadata and user bios' which may not be accurate. Expression of ideological preferences and moral appeals can be both subtle and highly subjective.  Methods leveraged to identify ideology and moral foundations, while state-of-the-art, may suffer from inconsistencies and biases given the task's inherent complexity. While our method to identify issue relevant tweets is highly scalable, less explicit references to issues in tweets may be missed by this method. Future work could test a supervised classifier that may more accurately identify all relevant tweets. While we made an effort to combine available lists of US political elites on Twitter, this list is not exhaustive.

\subsection*{Ethical considerations}

All data used in this study is publicly available \cite{chen2020tracking, mbfc2023politics}. The study was deemed exempt from review by the Institutional Review Board (IRB), as it relied solely on publicly available data. Our study also adheres to Twitter's terms of service \cite{twitter2022policy}. Tweet objects contain user information and this brings user anonymity into question. Users can restrict their tweets by switching to a private account or by deleting their tweets. Additionally, we preserve user anonymity during analysis by relying on user IDs instead of screen names. However, we acknowledge that many Twitter users may not be aware their data is used for research purposes ~\cite{fiesler2018twitter}. In this article, we present aggregated statistics to address further mitigate this concern. The authors declare that there are no competing interests.


\bibliographystyle{IEEEtran}
\bibliography{main}
\end{document}